# Variability in Grasp Type Distinction for Myoelectric Prosthesis Control Using a Non-Invasive Brain-Machine Interface

C. Piozin[(1)], L. Bouarroudj[(1)], J.Y. Audran [(2)], B. Lavrard[(3)], C. Simon[(3)], F. Waszak[(1)], S. Eskiizmirliler[(1)]

[(1)] Université Paris Cité, CNRS, Integrative Neuroscience and Cognition Center, F-75006 Paris, France; [(2)] Otto Bock France, Otto Bock, Courtaboeuf, France ; [(3)]Pôle Ambulatoire, Pôle appareillage du membre supérieur, Institut Robert Merle d'Aubigné, Valenton, France.

**Abstract**
Decoding multiple movements from the same limb using electroencephalographic (EEG) activity is a key challenge with applications for controlling prostheses in upper-limb amputees. This study investigates the classification of four hand movements to control a modified Myobock© prosthesis via EEG signals. We report results from three EEG recording sessions involving four amputees and twenty able-bodied subjects performing four grasp movements under three conditions: Motor Execution (ME), Motor Imagery (MI), and Motor Observation (MO). EEG preprocessing was followed by feature extraction using Common Spatial Patterns (CSP), Wavelet Decomposition (WD), and Riemannian Geometry. Various classification algorithms were applied to decode EEG signals, and a metric assessed pattern separability. We evaluated system performance across different electrode combinations and compared it to the original setup. Our results show that distinguishing movement from no movement achieved 100% accuracy, while classification between movements reached 70-90%. No significant differences were found between recording conditions in classification performance. Able-bodied participants outperformed amputees, but there were no significant differences in Motor Imagery. Performance did not improve across the sessions, and there was considerable variability in EEG pattern distinction. Reducing the number of electrodes by half led to only a 2% average accuracy drop. These results provide insights into developing wearable brain-machine interfaces, particularly for electrode optimization and training in grasp movement classification.

## Background

The loss of one or more limbs profoundly impacts the quality of life of amputees. This loss impairs their ability to carry out simple tasks, restricts their mobility, and necessitates the use of assistive devices. Among these devices, and for cases of upper limb amputations, myoelectric hand prostheses are the most commonly used to restore mobility. These prostheses operate by recording muscular activity via Electromyography (EMG) through surface electrodes placed on the stump of the missing limb. Typically, one or two surface EMG electrodes are used to control a few degrees of freedom, mainly opening and closing movements. Regardless of whether amputees use myoelectric prostheses or other types of devices such as passive or body-powered prostheses, research has shown that a substantial proportion—ranging from 10% to 50% of prosthesis users (depending on age)—discontinue using their devices [1]. The reported reasons for discontinuation include excessive fatigue, discomfort, and the heavy weight associated with the devices, making them unsuitable for daily use. This abandonment issue is particularly critical due to the psychological, functional, and social challenges already posed by amputation. The process of amputees adapting to a prosthesis is complex **Erreur ! Source du renvoi introuvable.** as they deal with limited motor control, reduced sensory feedback compared to their original limb, and the presence of phantom limb sensations.

To address those challenges, current research work has been dedicated to improving the control, quality, and performance of myoelectric prostheses along two axes of work. The first one is based on the use of Pattern Recognition techniques necessitating the increase of the number of EMG electrodes to characterize patterns of muscular activation when performing different movements [3]. Even though some devices using this technology are commercially available it is still difficult to evaluate the performance difference between using Pattern Recognition compared to conventional control [4]. Short training time and the small sample size of subjects are some of the main reasons behind the lack of clear results [5],[6]. The second axis is based on the use of the Targeted Muscle Reinnervation (TMR) procedure, which consists of rerouting several peripheral nerves on other muscles and using EMG electrodes on these new recording sites to control the prosthesis [7],[8]. This technique is particularly helpful for above-elbow amputees and has become over the years a standard procedure, yet because of the use of EMG electrodes, the TMR still suffers from the drawbacks of those electrodes [9]. While the EMG signal offers reliable and robust control over the device, the performance of the prostheses for both techniques is dependent on the quality of the muscles on the residual limb and the level of amputation, which can limit the number of movements that can be performed. If the quality of the EMG signals does not allow for satisfactory control over the prosthesis, or if no signal can be recorded, a solution is to use the source of movement intention in the cortex. Such a control system could be applied regardless of the level of amputation or limb absence, potentially targeting a wider range of patients.

The application of Brain-Machine Interfaces (BMIs) can therefore be considered as a new candidate to improve the control performance over prostheses by using brain activity as the control signal. BMIs are mainly composed of three interconnected parts, the Brain, Machine, and Interface, and can be divided into three distinct categories: invasive, semi-invasive, and non-invasive depending on the way of recording brain activity, focus will be put on non-invasive BMIs in this work. They rely on the recording of brain activity from an individual's scalp. The most commonly used technique is ElectroEncephaloGraphy (EEG) which refers to recording the brain's electrical activity using electrodes located all over the scalp. Studies showed that EEG-based BMI could allow the control of lower-limb exoskeletons [10], of a robotic arm in two dimensions [11] or a prosthesis with a steady-state visual evoked potentials BMI [12]. However, fewer studies are tackling the decoding of grasp movements from the same limb for the control of prostheses for upper-limb amputees. Indeed, most of the EEG-based studies rely on able-bodied subjects and the use of discriminative EEG patterns like Left/Right Motor Imagery. Yet, a shift in recent years saw the appearance of more studies recruiting amputees on top of able-bodied subjects. In particular, studies showed how amputees could control robotic hands via a BMI decoding movement kinematics (joint angle velocities for example) from EEG activity [13],[14]. Other studies are developing hybrid EMG-EEG BCI for the control of a prosthetic arm, which is an interesting and promising axis of research as both recording techniques could benefit from each other [15]-[17]. Nonetheless, there is a lack of research focused on controlling conventional myoelectric prostheses, despite their widespread use among the amputee population. (**) This means that findings from existing studies may not effectively benefit this population in the short term.

This paper presents the results of the second stage of our work aimed at assessing the feasibility of an EEG-based Brain-Machine Interface (BMI) designed for controlling a commercially available Myobock© myoelectric prosthesis. The evaluation starts with a basic synchronous paradigm involving a limited set of grip types for classification. Two pipelines are employed for feature extraction and off-line classification of EEG data recorded in three experimental sessions. The computed parameters are then utilized in a control test of the system. The first pipeline combines the classical Common Spatial Patterns technique with Wavelet Decomposition to compute features classified by four different classifiers. The second pipeline is based on Riemannian Geometry, incorporating two additional decoding algorithms. The paper introduces a new apparatus designed for healthy and amputee subjects to use the prosthesis during experiments. Given the inherent variability in EEG signals within and between sessions, the study explores changes in EEG signals associated with different movement intents and assesses the evolution of classification performance across the three sessions. The investigation extends to differences in offline classification performance between healthy and amputee subjects and for varying numbers of electrodes on the scalp.

## Methods

**Participants**

Twenty healthy participants (9 females and 11 males, 17 Right-handed, 3 Left-Handed, 30.05 ± 9.28 years old,) composed the control group and four amputated subjects (4 males, 2 transradial amputations of the left hand and 2 of the right hand, 60.25 ± 3.20 years old) participated in this second stage of the work. For the amputees, four subjects participated in the first session, three subjects in the second session (#1, #2 and #3), and two in the third session (#1 and #3). All participants were given information in oral and written forms before giving written informed consent to participate in the study. The protocol was performed in accordance with the Declaration of Helsinki and with the approval of the Comité de Protection des Personnes Ile de France II. The inclusion criteria for the amputated subjects were the following: persons above 18 years old with a transradial amputation and users of a myoelectric prosthesis. All amputees but Amputee #2 declared using their myoelectric prostheses - a polydigital hand (Michelangelo©, Ottobock) daily, and Amputee #2 is using his prosthesis (Myobock SensorSpeed©, Ottobock) occasionally.

**Experimental Protocol**

The experimental protocol which was followed by the able-bodied and amputee subjects during the experiments is separated into three distinct sessions. During each session, subjects perform or imagine a grasping sequence (**Figure 1**(A)) using three different objects (a cube, a cup, and a clothespin, respectively named BB, CUP, and CS) which are commonly used during dexterity tests to evaluate activities of daily living (ADL). They grasp, move, and release (GMR) the objects repeatedly, recording 30 trials with each of the three objects and 10 baseline/resting state trials. Recordings are conducted under three conditions: Motor Execution (ME), where they perform the movements, Motor Imagery (MI), where they imagine the sequence, and Motor Observation (MO), where they are moving the prosthesis towards the object with their arm but the opening and closing movements are controlled by the computer (similar to the observation protocol applied in [18],[19]). During the first session for the Motor Execution, the subjects perform the sequence with their daily prosthesis or their dominant hand (respectively for the amputees and the able-bodied). For the Motor Observation condition, in session 2 both groups of subjects are using the custom prosthesis and, in session 3, a sensory feedback system is added. The prosthesis is attached on top of the prosthesis socket for the amputees and a special apparatus is used with able-bodied subjects to allow them to use the prosthesis (see **Figure 1**(B)).

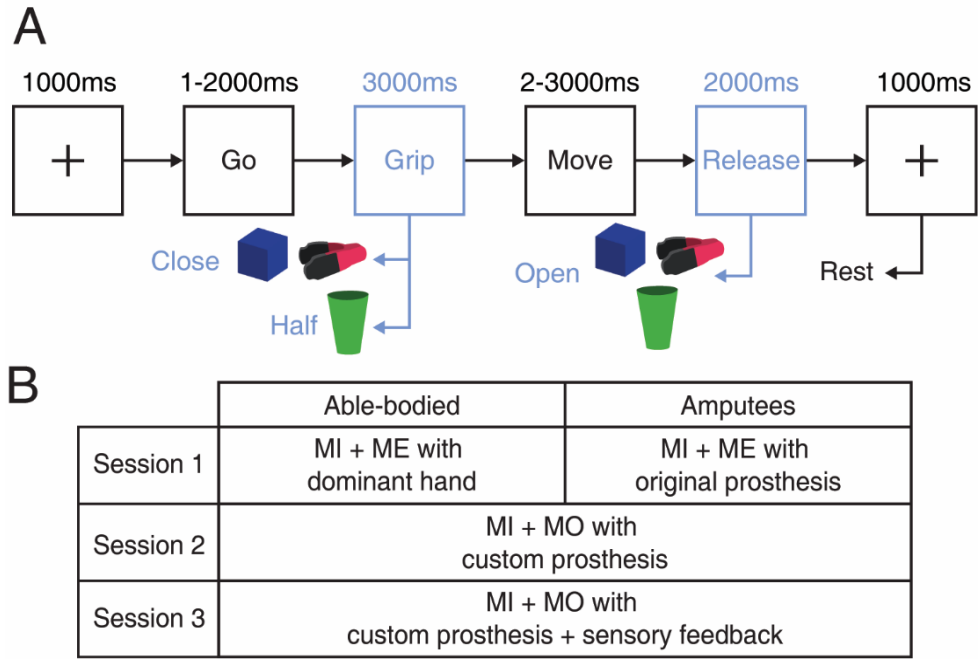

*Figure 1.* (A) Grasp-Move-Release (GMR) sequence performed by the amputees and able-bodied subjects during the experiments. The EEG signal of interest is obtained from the first second of the Grip and Release phases for the Tripod Grip (Grip phase of the cube and clothespin objects, referred to as BB and CS), Power Grip (Grip phase of the cup object, named CUP) and Open (Release phase for the 3 objects) movements. In contrast, the EEG signal for the Rest class is obtained from a fixation cross (resting-state trials). (B) Description of the three experimental sessions (S1, S2, S3) performed by the subjects.

**Experimental Setups**

During all experiments with able-bodied subjects, EEG signals were recorded at 2000Hz using a 64-electrodes ActiCHamp system (BrainProducts GmbH, Gilching, Germany). We used the PyCorder software during the signal acquisition process while the stimuli presentation was based on a Matlab script using the Psychophysics toolbox [20]. With amputees, EEG signals were recorded at 125Hz using a 16-electrodes OpenBCI system (OpenBCI, United States). The signal acquisition and stimuli presentation processes were done with a custom C++ program written and compiled in the QT IDE. The OpenBCI EEG recordin system was used with the amputees as they participated in the experiments outside of the laboratory, directly in the rehabilitation centre (IRMA) with recording sessions scheduled after meeting with the medical staff. To account for the differences regarding the number of electrodes between the two systems, the results presented will therefore mention when the same electrodes (in locations and number) were used to obtain the results.

As previously mentioned, both groups of subjects used a custom version of a Myobock prosthesis for sessions 2 and 3. The Myobock prosthesis has been modified so that the two directions of the single degree-of-freedom (DoF) for opening and closing are not controlled with EMG electrodes but through 2 Digital-Analog Converters (DACs) connected to an Arduino Nano microcontroller (**Figure 2**(C)). The DAC can send to the Myobock prosthesis an input voltage between 0 and 5V. Experimentally we noted a minimum of 0.6V to see the prosthesis move and a maximum of 1.6V for which there is no difference in the opening/closing speed beyond that limit. Similar results have been obtained by Nakamura and colleagues [21], noting a minimum input voltage of 0.56V and a maximum of 1.5V. It is worth highlighting that we were able to test with several different Myobock prostheses over the project, with left-handed or right-handed models and we found some small differences in the required voltages. Thanks to the DACs we can not only set the voltage to several values within the 0.6-1.6V interval but also define how long we are delivering this voltage. In other words, both the amplitude and the time are two parameters when setting a voltage on the DAC, much like what we observe with EMG electrodes where subjects control the intensity of the contraction and its duration. Therefore, we implemented in the microcontroller code a lookup table that defines the value (in mV) and the duration (in ms) of the voltage for a given movement. When a specific movement has to be triggered, the parameters are retrieved from this lookup table and sent to the corresponding DAC.

During the EEG recording sessions, the Myobock prosthesis is controlled by the stimuli presentation scripts on the computers; in the Grip and Release phases, commands are sent from the computers to the prosthesis one second after the beginning of the phases to respectively close (at two different aperture sizes for the tripod and precision grips) or open the prosthesis. Able-bodied subjects were able to use the prosthesis thanks to a 3D-printed apparatus. During session 3, a sensory feedback system, composed of force-sensing resistors and an arm bracelet equipped with vibrating motors communicating via Bluetooth with Bluno microcontrollers, was used to elicit vibrotactile feedback on the upper arm of the subjects (**Figure 2**(D)).

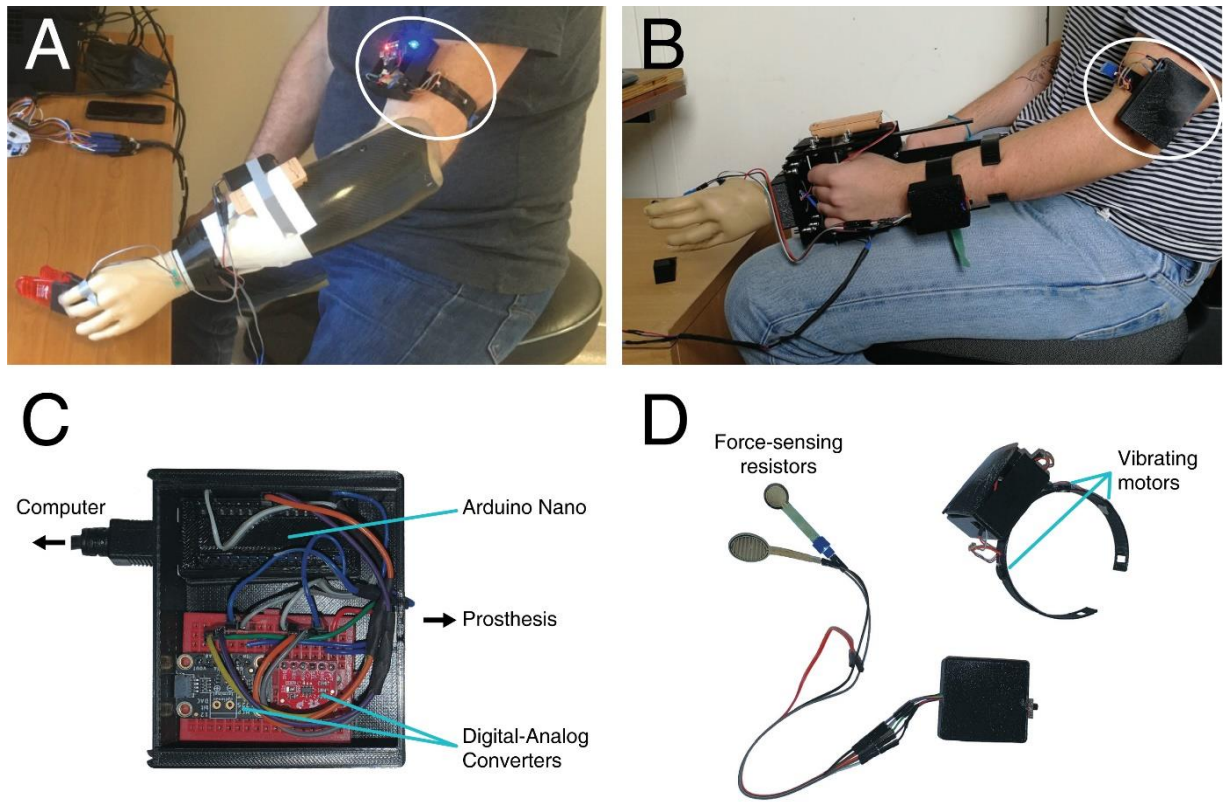

*Figure 2.* (A) Amputee subject using the custom Myobock with Sensory feedback (circled in white, here without top cover) device during the GMR recording with the CS object. (B) Able-bodied subject using the custom Myobock with the 3D-printed apparatus + sensory feedback (circle in white, with top cover) device during the GMR recording with the BB object. (C) 3D-printed enclosure with the electronic parts sending movement commands from the computer to the prosthesis. The value sent to the Arduino is translated into an input voltage to the corresponding DAC which then triggers a movement on the prosthesis. (D) Sensory feedback system. The first part contains force-sensing resistors (FSRs) that are located on the fingertips of the prosthesis (thumb and index+middle) while the second part contains vibrating motors that vibrate proportionally with the force information on the FSRs. Both parts communicate wirelessly thanks to Bluno microcontrollers located in the 3D-printed enclosures.

**Signal Processing and Feature Extraction**
*Preprocessing*

EEG datasets are first segmented using EEGlab [22] functions on Matlab to obtain the EEG signal of interest from the first second of the Grip and Release phases for the Tripod Grip (when the subjects perform the experiments with the BB and CS objects), Power Grip (with the CUP object) and Open (with the BB, CS, and CUP objects) movements (**Figure 1**(A)) while the data for the Rest class are obtained from the first second of the baseline/resting state trials. The 125 Hz EEG data recorded during the experiments with amputees were not downsampled while the EEG data recorded with able-bodied subjects were downsampled from 2000Hz down to 250Hz to reduce the size of the datasets. The EEG data are then notch filtered at 50Hz and bandpass filtered between 8 and 30Hz and a data augmentation technique (analogy method from [23]) is used to increase the trials from 30 to 90 for the Tripod Grip (TG), Power Grip (PG) and Open classes and from 10 to 90 for the Rest class to give an array of EEG amplitudes $X_k \in \Re^{N \times S}$ per electrode where $N$ is the number of trials (90), $S$ the number of samples in a trial (250 or 125, respectively for the able-bodied and the amputees) and $k$ is the class (movement types) (4). Finally, the whole data are normalized using the 50$^{th}$ percentile as the centering value and the difference between the values of the 95$^{th}$ and 5$^{th}$ percentiles as the scaling values computed by combining the $X_k$ of the four classes for each electrode (63 for the able-bodied and 16 for the amputees):

$$X'_k = \frac{X_k - q_{50}}{q_{95} - q_{50}} \ for \ k = 1, \dots, 4 \quad (1)$$

*Feature Extraction*

A 5-fold cross-validation approach was employed to partition the pre-processed data into 5 groups, with 4 groups dedicated to training/fitting the model and 1 group to evaluate the model; the procedure is then repeated so that each of the 5 groups is used to evaluate and to train the models. Consequently, the performance metrics presented herein represent the mean performance across these 5-fold iterations. The feature extraction pipeline is based on a combination of Common Spatial Patterns (CSP) and Wavelet Decomposition (WD). With the CSP technique, we applied shrinkage regularization to estimate the covariance matrices in a more representative and reliable way (3), as detailed in [23]:

$$C_k = \frac{1}{N} \sum_{j=1}^{N} X'_{j,k} (X'_{j,k})^T \quad (2)$$

$$\widetilde{C_k} = C_k + \lambda I \ \ for \ k = 1, \dots, 4 \quad (3)$$

where $N$ is the number of trials (90), $X_{j,k} \in \Re^{E \times S}$ is the $j^{th}$ EEG trial from class $k$ with $E$ the number of electrodes and $S$ the number of samples, $T$ denotes transpose, $I$ is the identity matrix and $\lambda$ is a free parameter whose optimal value is computed with Ledoit & Wolf shrinkage [24].

As the CSP algorithm is commonly used for binary classifications, we computed the CSP filters by comparing the 4 types of movements on a one-vs-one basis, leading to the computation of 6 distinct CSP filters. With the WD method, we used coefficients obtained from the last level of decomposition, set at 2 or 3 respectively for the OpenBCI or ActiCHamp systems. Subsequently, the log variances of the Wavelet Decomposition feature vectors were utilized as inputs for the classification algorithms, leading to inputs $Y \in \Re^{F \times 2N}$, where $F$ corresponds to the number of CSP features kept after CSP (12 or 16 respectively for the able-bodied and amputees) and $2N$ corresponds to the number of trials (180, 90 for each class), for each of the 6 binary models.

In recent works [25]-[27] studying the decoding of the EEG data, the use of Riemannian Geometry for Brain-Machine Interface has been reported to yield good results for Left/Right movement classification but it hasn't been employed for the classification of movements from the same limb. Moreover, the use of learning algorithms based on Riemannian Geometry requires another feature extraction method, slightly different from what we presented in the beginning of this section. Starting from the same pre-processed data as the first pipeline, the initial step involved computing covariance matrices for all trials across the four classes of movements. Similar to the CSP approach mentioned earlier, the covariance matrices were estimated using the same shrinkage method. The subsequent step involved computing the mean covariance matrices for each class, achieved with the assistance of the PyRiemann toolbox [28]. This toolbox offers various techniques for computing centroids, and we selected the Log Euclid technique after conducting trials with different methods, including Riemannian mean, arithmetic mean, and harmonic mean.

The final step within the Riemannian geometry pipeline entailed projecting the covariance matrices onto the tangent space of the two classes:

$$S_i^j = Log\left(M_i^{-\frac{1}{2}} C_i^j M_i^{-\frac{1}{2}}\right) for \begin{cases} i = 1,2 \\ j = 1,90 \end{cases} \quad (4)$$

where $M_i$ is the geometric mean for class $i$, $C_i^j$ the covariance matrix for class $i$ and trial $j$, and $Log()$ is the matrix logarithm function. The rationale behind this projection is succinctly explained in [29]. In essence, the tangent space of the Riemannian space is Euclidean, which enables the use of conventional machine learning algorithms. The feature vectors consist of vectorized projected covariance matrices:

$$z_i^j = uvec(S_i^j \circ Q) \quad (5)$$

where $Q$ is a matrix whose elements are 1 on the diagonal and $\sqrt{2}$ elsewhere, ∘ is the element-wise product and $uvec()$ vectorizes the upper triangular part of the product.

*Movement type prediction*

As mentioned at the beginning of section 2. the role of the movement type prediction is twofold: to evaluate how representative the inputs obtained by the feature extraction method are, and to provide repeatable, generalizable, and sufficiently robust output commands for the desired types of GMR movements (four in our cases: Open, Tripod Grip (TG), Power Grip (PG), Rest). This crucial role of the classification module of the interface part of our BMI system therefore led us to test classically used but also recently proposed algorithms which required the use of distinct feature extraction pipelines for different algorithms. More specifically, the classification performance of the following classifiers: An Artificial Neural Network (ANN) composed of a 15-neuron Hidden Layer with scaled conjugate gradient backpropagation, two SVMs with Linear and Radial Basis Function kernels, and Linear Discriminant Analysis (LDA) have been assessed with the first feature extraction pipeline combining the Common Spatial Pattern and Wavelet Decomposition. In addition, two other classifiers have been tested with the Riemannian geometry-based feature extraction. The first classifier, Minimal Distance to Mean (MDM), utilizes mean covariance matrices from the two classes as inputs and calculates the distance between these means and a trial to be labeled. In this case, the distance metric employed is the log-euclidean distance between the mean covariance matrices and the covariance matrix of the trial. The second classifier, Tangent Space SVM (TS-SVM) with an RBF kernel, takes feature vectors obtained after projection onto the tangent space as inputs. We computed the mean Accuracy, F1-Score, and Precision over the 5-fold cross-validation for each of these classifiers.

*Effect of the combination of electrodes on prediction performance*

We analysed offline the data of the 20 subjects in the able-bodied group with 64-electrodes recordings to look into the potential reduction of the number of electrodes for the BMI system as well as investigate the role of other brain areas for the decoding of motor intention. As shown in **Figure 3**, 9 different combinations of electrodes were investigated.

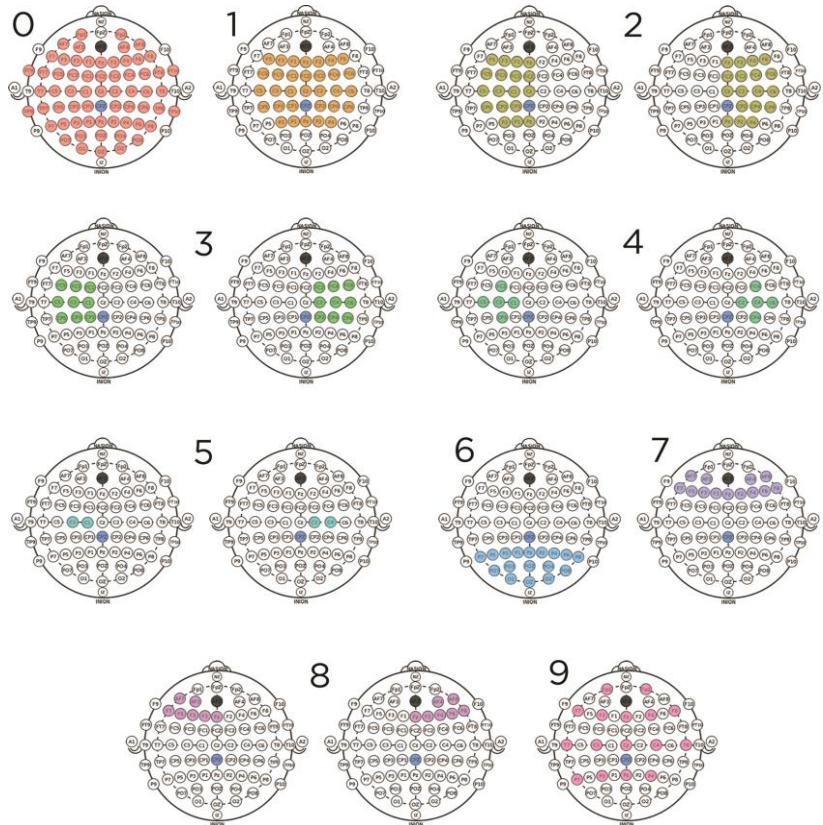

*Figure 3.* Combinations of electrodes used to study how the prediction performance is impacted when reducing the number of electrodes. Combination 0 (64 electrodes) is the set of electrodes that was used during the EEG recordings with the able-bodied. Combination 1 is composed of 32 electrodes centered around the motor cortex. Combinations 2, 3, 4, and 5 gradually reduce the number of electrodes around electrode C3 for the right-handed subjects and around electrode C4 for left-handed subjects, corresponding to the contralateral part of the cortex involved in motor functions. Combination 6 contains electrodes localized around the Occipital cortex, which is related to vision. Combinations 7 and 8 are electrodes over the frontal lobe, with combination 8 presenting the same contralateral splitting as combinations 2-5. Finally, combination 9 corresponds to the locations of the 16 electrodes on the OpenBCI system used with the amputees.

## Results

### *Differences between binary classification models on performance*

As a first step, we investigated which movements are more distinguishable from the others based on the offline classification performance of the six binary classification models. **Figure 4**.(B) compiles the results from session 2 conducted with able-bodied subjects. The analysis reveals performances above chance level for all the pairs (56.1% computed with p=0.05 significance threshold, based on [30]) no discernible differences between the TG/Rest, PG/Rest, and Open/Rest pairs. Additionally, there are no significant differences between the TG/PG and PG/Open and the TG/Open and PG/Open pairs. Notably, regardless of the condition, binary classification models comparing a movement (TG, PG, or Open) against Rest consistently demonstrate significantly higher performances compared to other models. Conversely, performances from the TG/Open model are consistently lower than those of other models.

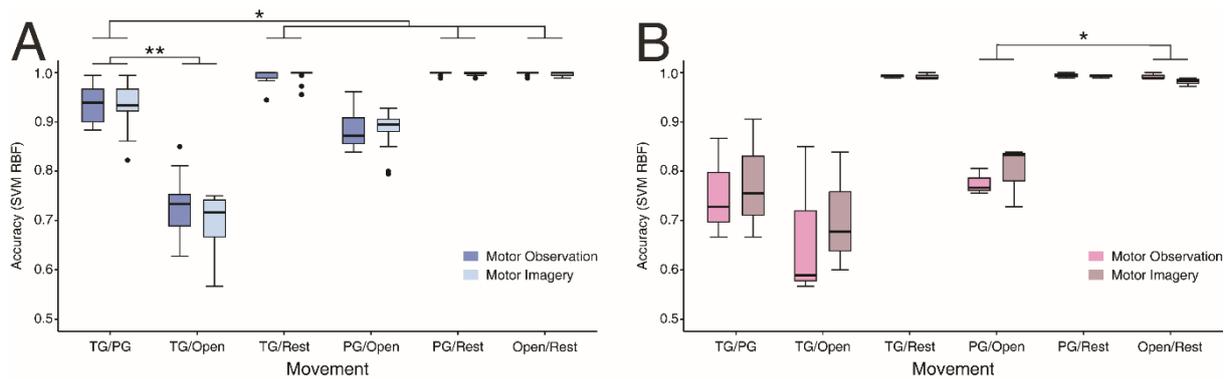

*Figure 4.* Classification Accuracy for the Support Vector Machine with Radial Basis Function kernel for each of the binary models during Session 2 in the two conditions (A) for the able-bodied group. (B) for the amputee group. The * and ** indicate significant differences with p<.05 and p<.01 respectively. For simplification purposes in (B), significant differences between TG/Rest and PG/Open and between PG/Rest and PG/Open were not represented.

The results obtained with the amputee subjects for session 2 are presented in Figure 4 (B). They are similar to those obtained with the able-bodied, with the exception that there is now no difference between TG/PG, TG/Open, and PG/Open pairs.

Furthermore, we investigated whether there were any differences between the three conditions during the three sessions in the two groups of subjects. For this, we performed a Wilcoxon test on each algorithm and binary model between the performances of the two conditions. When using the SVM RBF algorithm, we observed that there are no significant differences in the performances between the execution, the observation, and the imagination of the movements over the three sessions for both groups.

### *Impact of Sensorimotor cortex excitability (SCE) conditions on classification performances between groups*

We analysed the impact of different sensorimotor cortex excitability conditions on the control performances of two subject groups. To have the same number of EEG signals for each group coming from the same regions of the cortex, for able-bodied participants we selected 16 electrodes on the ActiCHamp system (having 64 electrodes) matching with the locations of the 16 electrodes on the OpenBCI system used with amputee subjects (see combination 9 in **Figure 3**) and then processed only those EEG signals. These comparisons were conducted for each session, under the three conditions (ME, MO, and MI), and across the six binary classification models. The significance of these comparisons was evaluated through Wilcoxon tests. As represented in **Figure 5** in both the Motor Execution and Motor Imagery conditions for session 2, we observe no significant differences in the offline performances between the two groups for the six binary models when using the SVM RBF classifiers. However, in Motor Observation even though there are no significant differences, the performances of the amputees were lower than that of the able-bodied ones (see TG/PG, TG/Open, and PG/Open in **Figure 5**(A),left). In Motor Imagery (**Figure 5**(A), right) we observe that the performances of the amputees are closer to those of the able-bodied (see TG/Open and PG/Open). Still, these results can be nuanced by the low number of subjects in the amputee group (n=4 for S1, n=3 for S2, and n=1 for S3). To act on this issue, we applied a bootstrap procedure with 1000 repetitions of the Wilcoxon test. For example, for session 2, we randomly selected three able-bodied subjects among the 20 and compared them with the three amputees who did the second session. After the bootstrap, in Motor Observation and Motor Imagery, we obtained no significant differences in the performances for all the binary models compared to when the Wilcoxon tests were done with the full number of able-bodied subjects.

When comparing the performances of the two groups of subjects this time with the 64 electrodes with the Able-bodied (**Figure 5**(B)) we observe some significant differences for the TG/PG ($p < .01$ and $p <$

.05, respectively for MO and MI), PG/Open ($p < .01$ and $p < .05$, respectively for MO and MI) and Open/Rest ($p < .05$ and $p < .01$, respectively for MO and MI). It is interesting to note that even though there are differences obtained for the Open/Rest binary model, the performances for both groups are around 100% accuracy.

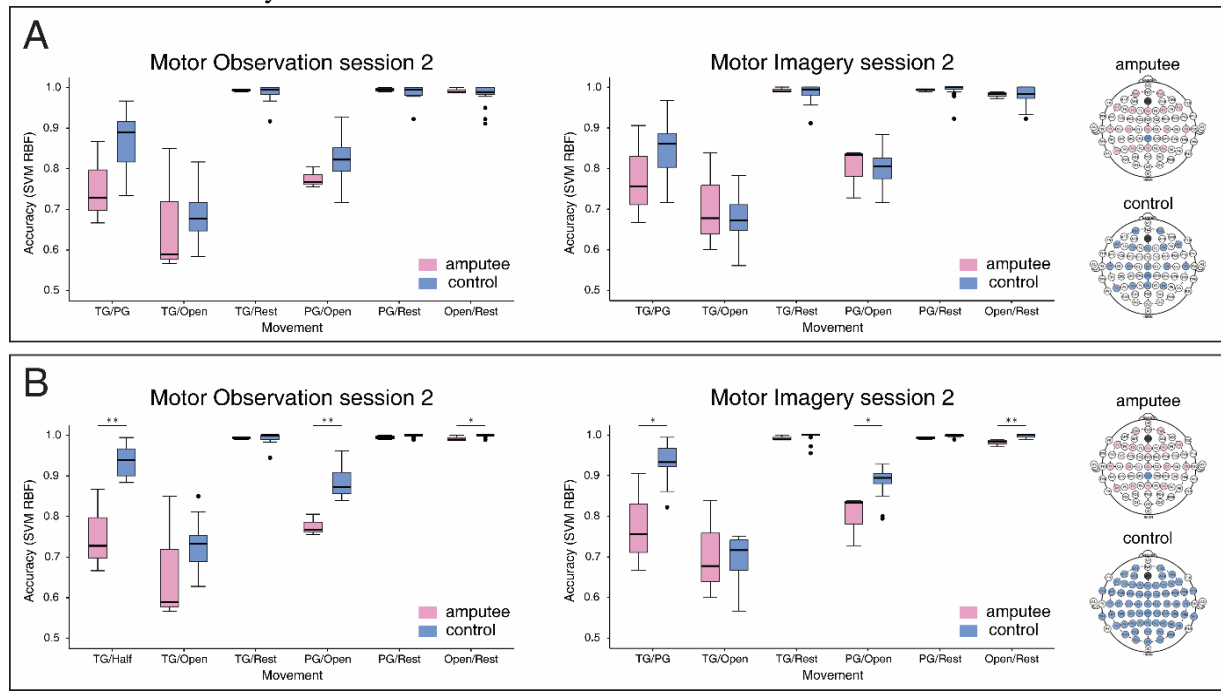

*Figure 5.* Classification Accuracies for the Support Vector Machine with Radial Basis Function kernel for each of the binary models for the two groups of subjects in the Motor Observation (left) and Motor Imagery (right) conditions of session 2. (A) The performances are obtained after matching the electrodes between the two groups. (B) The performances are based on a comparison of the 64 electrodes for the control group with the 16 electrodes of the amputee group.

*Evolution of the control performance across experimental sessions*

Here, we first studied whether the performances improved or remained stable across the three sessions. From **Figure 6**(A) we can see the performance of each of the binary models is stable along the sessions; this has been confirmed with Wilcoxon tests between the sessions which revealed non-significant differences for the six models and the three conditions.

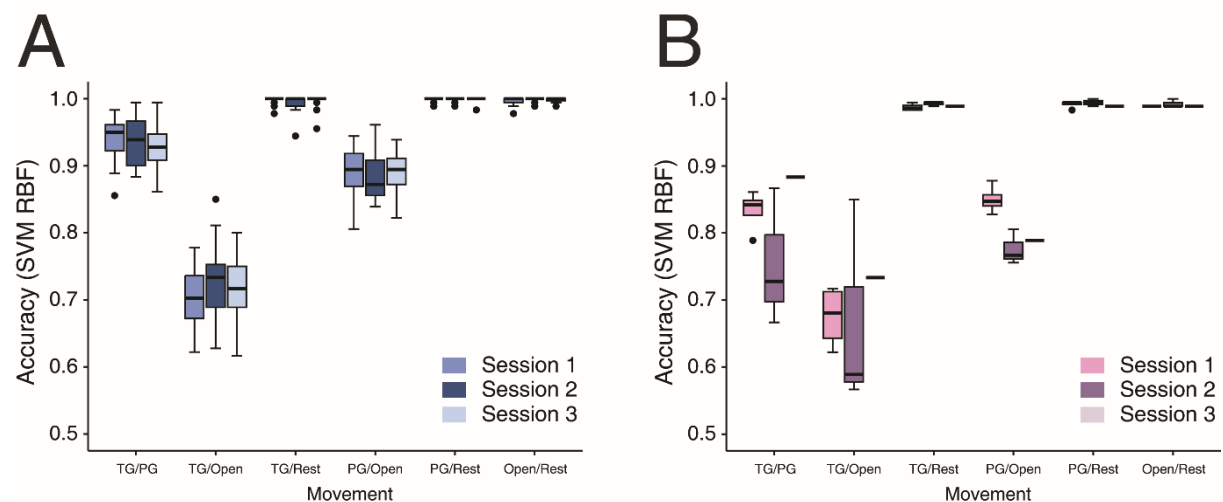

*Figure 6.* Accuracies for the 6 binary models along the three sessions in the Motor Execution (S1) and Motor Observation (S2 & S3) conditions. (A) Accuracies of the RBF SVM in the able-bodied group. (B) Accuracies of the RBF SVM in the amputee group.

Even though the small number of amputee subjects (n=4 for S1, n=3 for S2, and n= 1 for S3) only allows us to look at a trend in the evolution of performance, as displayed in **Figure 6**(B), it is interesting to examine the differences between Session 1 and Session 2, as well as between Session 2 and Session 3. Regarding the first comparison (S1 vs S2), we observe a drop in the performance for the binary classification models TG/PG, TG/Open and PG/Open. For the difference between Session 2 and Session 3, it is impossible to conclude as only one subject was able to participate in the third session. We note however that the classification performance increases for the models TG/PG, TG/Open, and PG/Open.

*Evolution of the distinction of movements across experimental sessions*

In a subsequent phase, we calculated the class distinctiveness (classDis), a performance metric introduced by Lotte and Jeunet [31]. This metric assesses the separation between two EEG patterns generated by a BMI user by examining the distance between the means of each class and their dispersion around those means. Our analysis focused on observing the evolution of this metric across the three sessions for both subject groups and under each condition. The ideal outcome would be an improvement in distinctiveness over time, reflecting enhanced subject training for the task.

As seen in **Figure 7** for the able-bodied group in Motor Execution, we can observe that the distinctiveness is slightly improving over time for some of the binary models. In particular, if we try to fit a line over each models the distinctiveness of PG/Open, PG/Rest, and Open/Rest is close to following a linear evolution. At the same time, we do not observe the same evolution for TG/PG, TG/Open, and TG/Rest as the distinctiveness does not seem to improve over time. After running Wilcoxon tests to compare between sessions, we found non-significant differences except in Motor Execution/Observation with an improvement in TG/Rest between the first and third sessions, and in Motor Imagery where the classDis is better in the second session compared to the first for TG/Rest, PG/Rest and Open/Rest.

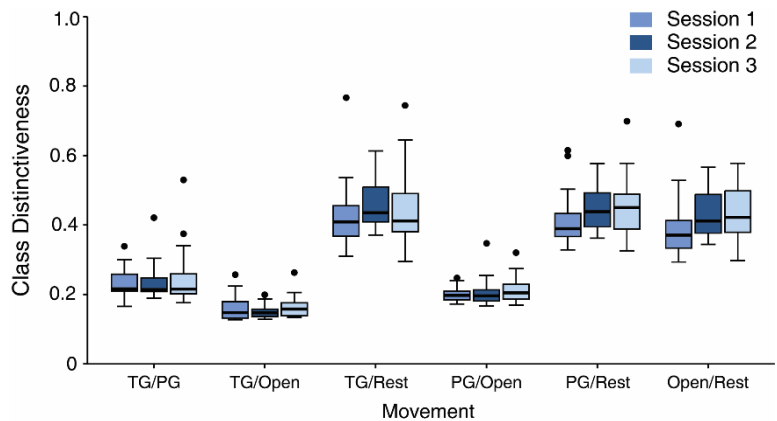

*Figure 7*. Evolution of the class distinctiveness metric along the three sessions in the Motor Execution (S1) and Motor Observation (S2 & S3) conditions in the able-bodied group.

We also looked at individual results for able-bodied subjects whose distinctiveness either increased, remained constant or decreased across the three sessions to see if the evolution in class distinctiveness is translated into the decoding performance.

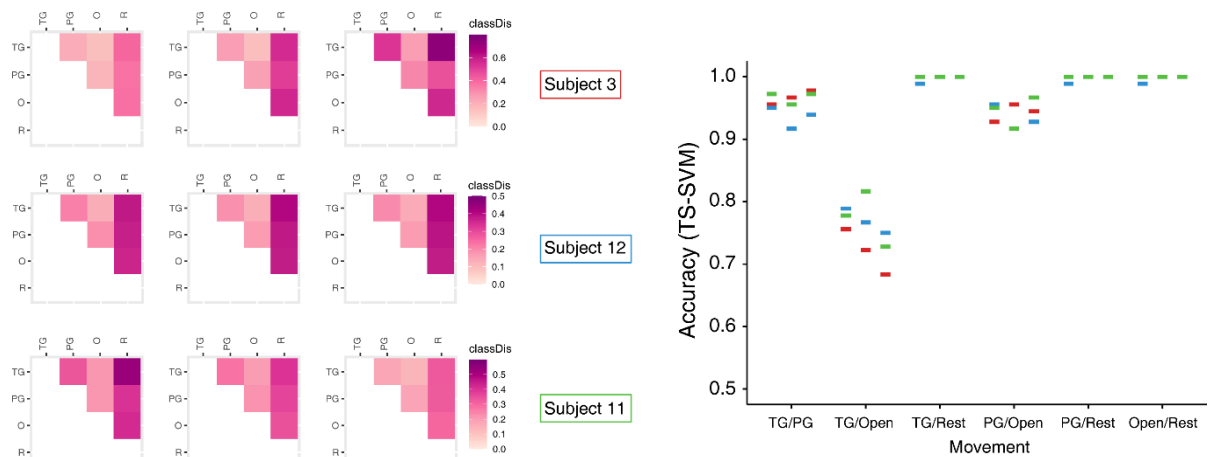

*Figure 8.* (A) class distinctiveness between each pair of movements for the three experimental sessions (ME, MO, MO + sensory feedback) for three subjects in the able-bodied group. (B) Accuracy of the TS-SVM for the three experimental sessions. Each color represents one of the three subjects from the able-bodied group.

From **Figure 8**, over the length of the experiment, we can see that the distinctiveness is increasing for subject 3, remaining constant for subject 12, and decreasing for subject 11. If we compare that with the decoding performances, we observe several cases. First, the performance is increasing with the distinctiveness. This is particularly visible for subject 3 with the TG/PG model where we observe an increase in both metrics. We can also note that even though the class distinctiveness of a movement (TG, PG, or Open) with Rest gets higher, this does not necessarily lead to an increase of the decoding performances as they are already reaching 100% accuracy. Second, the performance decreases while the distinctiveness increases. Indeed, still for subject 3, we see that the distinctiveness between Tripod and Open is slightly increasing while the performance for the corresponding model is dropping from around 76% to 68% accuracy. Third, the performance maintains or increases while the distinctiveness decreases. This trend is visible for Subject 11 with the distinction between TG/PG where we can observe a drop in distinctiveness which is not affecting the decoding performance as the decoding performance between the first and third session stays around 98%.

We obtain different profiles of class distinctiveness and decoding performances when looking at the Motor Imagery data for the three sessions.

### *Effect of electrode selection on offline performance and separability*

The EEG data collected from twenty healthy subjects who used the 64-electrodes ActiCHamp EEG helmet during the experiments was processed for eight different combinations of the electrodes shown in Figure 3 and the performance metrics were compared across the experimental sessions.

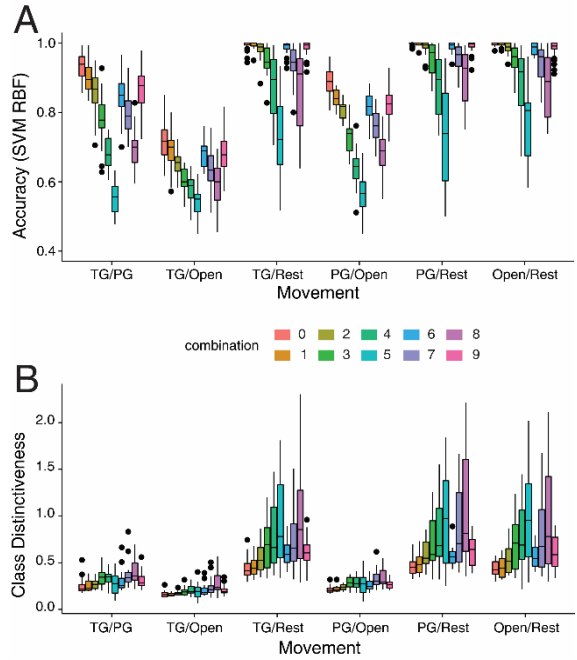

*Figure 9.* Performance metrics for the six one-vs-one models for the ten combinations of electrodes for the third session in Motor Observation for able-bodied subjects. (A) Classification performance for the RBF SVM. (B) Class distinctiveness.

**Figure 9**(A) shows the results of the electrode selection for the third session. We can note that the performances of the various test combinations are all lower than the performance of combination 0, especially for the TG/PG, TG/Open, and PG/Open binary models. Wilcoxon tests between each of the 9 combinations with the original one all gave significant p-values with significantly lower performances, except for TG/Rest and PG/Rest of combination 1 for the three sessions. As shown in Figure 7., Lowest performances were obtained from combination 5 while performances from the other combinations evolved in the same range. Similar results were obtained for the first and second sessions. We also computed the corresponding loss of accuracy obtained when selecting one of the nine combinations compared to the original configuration with all of the 64 electrodes. For the SVM RBF, the drop in accuracy ranges from 2% for combination 1 to 30% for combination 5.

We then looked at the changes in the class distinctiveness between combination 0 and the nine other subsets of electrodes. We observe from **Figure 9**(B) that for the binary models TG/Rest, PG/Rest, and Open/Rest the value of class distinctiveness is higher than the three other models where the changes in class distinctiveness for different combinations are less visible. Wilcoxon tests comparing combinations 1-9 to combination 0 for the six models in Motor Observation from Session 3 showed a positive significant effect of the electrode removal on the class distinctiveness except for combination 1.

## Discussion
### Differences between grasping movements and SCE conditions

Our study corroborates observations from [32], suggesting that the use of a binary classification model estimating the control command by comparing a movement (an active state of the prosthesis: Tripod Grip, Precision Grip, or Close) to the absence of movement (Rest) leads to higher accuracy. Interestingly, the offline performance tends to be lower when we try to estimate them using the binary model comparing antagonistic movements (e.g., Opening and closing the hand), the performance of around 70% for the comparison between Tripod Grip and Open is in line with [33] where they obtained similar performances for the classification of open and close motor imagery. However, the higher performance obtained with the binary model comparing Grip and Open differ from those observations, we hypothesize that task-related information from the cortex will have a role in the decodability between different patterns of EEG activity.

Furthermore, our results also showed no significant differences in offline performance between Motor Execution, Motor Imagery, and a hybrid of the two we referred to as Motor Observation. As such, our findings confirm previous studies that showed similar activation of the sensorimotor cortex during motor execution and motor imagery of hand movements with able-bodied subjects [34]. Moreover, even though it was reported that different cortical networks are active during executed and imagined hand movements of amputees [35], such differences are not seen in offline performances.

**Differences between amputees and able-bodied**

Our analyses showed that there were no significant differences when comparing amputees and able-bodied with different EEG systems in Motor Execution and Motor Imagery. However, while the results showed similar performances between the two groups with imagined hand movements, we still observed higher performances for the able-bodied in Motor Execution. This finding agrees with what was reported in [36] with EMG activity where participants with upper-limb amputations generated less separable patterns than able-bodied participants. However, as we used different EEG systems with the two groups, we cannot rule out the possible influence of a higher-quality system on the performance. Future studies may elucidate the influence of the recording system on the classification performance. Nonetheless, these results confirm and extend previous findings that showed that the use of a consumer-grade EEG system can compete with research-grade amplifiers, giving lower performances yet close and reliable EEG readings [37].

**Evolution of offline performance**

We found that there is no positive evolution of the offline performances through the three experimental sessions for the able-bodied group. On one hand, we did not expect differences in the Motor Imagery condition as the three sessions consisted of the same imagined movements. Therefore, apart from a potential learning effect of the task, there was no reason to expect a change. On the other hand, it is quite interesting to note that the introduction of the new device with the modified prosthesis (for the second session in Motor Observation) and then the sensory feedback device (for the third session in Motor Observation does not induce brutal changes in the performance of the BMI system compared to the first session where subjects are using their dominant hand to perform the movements. This finding is quite surprising as we expected to see a drop in the classification accuracy between the first and the subsequent sessions and to the best of our knowledge, we did not find any examples of similar experimental protocol in the literature. However, we believe that the custom apparatus is acting as a tool for the participants, potentially extending their hands and becoming embodied [38].

It is difficult to conclude regarding performance differences between sessions with the amputees because of the low number of subjects. As with the able-bodied, we hypothesized that the introduction in Session 2 of the custom Myobock to the subjects would lead to a drop in the offline classification, which is what we observed for the binary models comparing one movement with another. Regarding the difference between Session 2 and Session 3, corresponding to the addition of the Sensory Feedback system for the amputees, not only does this addition possibly increase the classification performance compared to Session 2 when comparing the movements with each other (and with no change when comparing a movement with Rest) but we can also note that the subject that did the third session obtained better performances compared to the first session. Literature suggests that the use of vibrotactile feedback increases accuracy in grip force tests [39], functional assessments [40], and motor tasks [41]for upper-limb amputees. Notably, the participant in our study who tested the sensory feedback system reported not having to rely on vision as much as without the feedback. Future studies will further investigate the role of sensory feedback on classification performance.

**Correlation between EEG pattern separability and offline performance**

We found that there is not a strong correlation between the separability of EEG patterns related to hand movements and offline performances. Our findings are in line with Franzke et al. [42] where they argued that higher performance comes with higher separability, but a higher separability does not necessarily mean higher performance. Indeed, our results suggest a large variability in the predictive power of the class distinctiveness metric, depending not only on the subject but also on the movements that were compared. That is, the subjects are getting better at creating distinguishable patterns between a movement (any active state of the prosthesis) and the absence of movement/Rest but this does not apply when the subjects try to create different patterns for two active states of the prosthesis. It is worth reminding that the sources of the activity for the TG, PG, and Open movements of the hand are localized in a very narrow region of the cerebral cortex as can be observed on the somatotopic maps [43], implying that higher separability is unlikely to happen or at the cost of longer training time. As reported in the literature, there is evidence of reorganization of the sensorimotor cortex following an upper-limb amputation [44],[45] which would potentially make the distinction of hand movements more difficult for amputees. The relation between separability and performance should be investigated in longitudinal studies.

**Use of a limited number of electrodes**

In an ideal setting, we would want to improve the accuracy of the system while also improving the distinction between the movements. However, as discussed in the literature [31],[42] the correlation between classification performance and separability of patterns is not linear. In our study, we extend those findings as we observe that if we select fewer electrodes there is a trade-off in that we need to sacrifice some of the accuracies if we want to have a better distinction.

Interestingly, results from the electrode selection process revealed a 2% drop in accuracy (averaged across the six binary models) when splitting the number of electrodes by 2, reducing the number of electrodes to 32. Furthermore, by splitting the number by 4 with combinations 2 or 9, we observed a loss of accuracy of about 3 to 5% compared to the original accuracy. As seen in **Erreur ! Source du renvoi introuvable.**(B), the more electrodes are removed the better the distinction between movements becomes. An explanation for this behaviour might be the focus on the sensorimotor cortex with the removal of electrodes that are less relevant for the distinction of movements in the frontal or temporal regions. The results obtained with combinations 6 and 7/8, respectively targeting the parietal/occipital and prefrontal cortex, also show an increase in the distinction of movements. These findings are supported by the literature which showed a contribution during visuomotor and hand movement tasks of 1) the parieto-occipital cortex [46],[47] and 2) the prefrontal cortex [48],[49]. Therefore, we hypothesize that covering a large area of the cortex while limiting the number of electrodes as depicted with combination 9 can lead to better separability and above chance-level accuracy for movement decoding tasks.

**Conclusion**

This study investigated changes in the distinction of four hand movement types performed by able-bodied participants and trans-radial amputees during three experimental sessions. We showed no significant differences between the two groups under equivalent conditions. Moreover, as we observed no differences between the Motor Imagery, Motor Execution, and Motor Observation conditions, it would be interesting to extend the study to the paraplegic population or, to some extent, to the stroke population. We showed offline performances above chance level in the two groups for the distinction of hand movements, and close to 100% when comparing movements with the absence of movement. We did not observe a positive evolution of the offline performance across the three sessions for the able-bodied, and we hypothesize a potential improvement using sensory feedback with the amputees which is to be investigated with more participants. Results from the separability of patterns suggest an improvement of this metric over time but with no correlation to the classification performance. Furthermore, an improvement in the distinction between movements is obtained by reducing the number of electrodes used, which has implications for the development of an acceptable and wearable system for amputees. The reduction of electrodes comes with a reduction in accuracy which can be limited by selecting relevant electrode locations. Further studies from the group will look into confirming the offline results with real-time tests and a larger sample size for participants with upper-limb amputations. Moreover, the performance of the BMI system will be assessed with functional tests with the amputees, comparing the performance with their myoelectric prostheses.